\documentclass[sigconf, nonacm]{acmart}
\AtBeginDocument{%
  }

\usepackage{enumitem,kantlipsum}

\usepackage{booktabs}
\usepackage{tabularx}
\usepackage{makecell}
\usepackage{tcolorbox}
\usepackage[table]{xcolor}
\newcommand{\textbox}[1]{%
  \noindent
  \begin{tcolorbox}[
    colback=gray!15,
    colframe=black,
    arc=2mm,
    boxrule=1pt,
    left=1mm,
    right=1mm,
    top=1mm,
    bottom=1mm,
    enlarge top by = 1mm,
    after=\vspace{0mm},
    valign=center,
    parbox=false
  ]
  #1
  \end{tcolorbox}%
}

\begin{document}

\title{How Do Agentic AI Systems Deal With Software Energy Concerns? A Pull Request-Based Study}


\author{Tanjum Motin Mitul}
\authornote{Both authors contributed equally to this research.}
\affiliation{%
  \institution{SQM Research Lab\\ Computer Science\\ University of Manitoba}
  \country{Winnipeg, Canada}
}

\author{Md. Masud Mazumder}
\authornotemark[1]
\affiliation{%
  \institution{SQM Research Lab\\ Computer Science\\ University of Manitoba}
  \country{Winnipeg, Canada}
}

\author{Md Nahidul Islam Opu}
\affiliation{%
  \institution{SQM Research Lab\\ Computer Science\\ University of Manitoba}
  \country{Winnipeg, Canada}
}

\author{Shaiful Chowdhury}
\affiliation{%
  \institution{SQM Research Lab\\ Computer Science\\ University of Manitoba}
  \country{Winnipeg, Canada}
}


\begin{abstract}
As Software Engineering enters its new era (SE 3.0), AI coding agents increasingly automate software development workflows. However, it remains unclear how exactly these agents recognize and address software energy concerns—an issue growing in importance due to large-scale data centers, energy-hungry language models, and battery-constrained devices. In this paper, we examined the energy awareness of agent-authored pull requests (PRs) using a publicly available dataset. We identified 216 energy-explicit PRs and conducted a thematic analysis, deriving a taxonomy of energy-aware work. Our further analysis of the applied optimization techniques shows that most align with established research recommendations. Although building and running these agents is highly energy-intensive, encouragingly, the results indicate that they exhibit energy awareness when generating software artifacts. However, optimization-related PRs are accepted less frequently than others, largely due to their negative impact on maintainability. 

\end{abstract}
\balance




\maketitle

\section{Introduction}
Energy consumption has become a core concern in software engineering due to its environmental, economic, and battery-life impacts~\cite{lee2024survey}. Data centers consume a substantial amount of electricity, and the growing adoption of Large Language Models (LLMs) further increases compute demand and energy use, thereby raising operational costs and carbon emissions~\cite{koot2021usage, faiz2023llmcarbon, argerich2024measuring}. On the user side, energy-inefficient software shortens battery life in mobile and IoT devices, influencing device preferences~\cite{pang2015programmers, user-care-battery,moura2015mining}.



In response, researchers and developers have long worked to reduce energy consumption at both the hardware and software levels~\cite{chowdhury2019greenbundle, osta2019energy, carroll2010analysis}. This has become even more important as in the SE 3.0 era, AI systems frequently produce and submit code changes as standard artifacts~\cite{li2025aidev, hassan2024towards}, many times as GitHub pull requests (PRs). Therefore, the ability of these agents to recognize energy issues can influence real-world energy use at scale~\cite{cursaru2024controlled}. Recent studies have compared the energy efficiency of AI-generated code and human-generated code, finding that AI-generated code can often reduce energy consumption~\cite{vartziotis2024learn, stivala2024investigating}. However, it remains unclear how these AI agents explicitly address energy consumption and which techniques they employ to improve software energy efficiency. This understanding is crucial for training future coding agents to make them more energy-aware. This study addresses this research gap by investigating the following two research questions using the AIDev~\cite{li2025aidev} dataset.

\textbf{RQ1:} \textbf{What aspects of energy concerns are considered by AI agents?}
After a rigorous manual annotation process, we identified 216 energy-explicit pull requests (PRs). 
We applied thematic analysis~\cite{CruzesD.S.2011RSfT} on these PRs and developed a five-category taxonomy: \textit{Insight}, \textit{Setup}, \textit{Optimization}, \textit{Trade-off}, and \textit{Maintenance}.

\textbf{RQ2:} \textbf{Which energy optimization techniques are followed by the AI agents?}
We extracted 21 optimization techniques employed by the coding agents in these pull requests and found that nearly all of these techniques align with research-based recommendations, suggesting that AI agents are well aware of effective energy optimization strategies. However, despite their potential, this category of PRs is accepted less frequently due to their impact on code maintainability, thus implying the importance of future research producing techniques that can make software artifacts (e.g., code) that are not only energy efficient but also maintainable.




To enable replication, we share our data and code publicly.\footnote{\url{https://github.com/SQMLab/LLM-energy}}
\section{Related Works}

Prior work has extensively studied software energy concerns and optimization techniques. Cruz and Abreu~\cite{cruz2019catalog} cataloged energy patterns by mining commits, issues, and PRs across Android and iOS apps, while Bao et al.~\cite{bao2016android} identified recurring power-management activities through repository mining. Li and Halfond~\cite{li2014investigation} empirically evaluated common energy-saving practices, showing that some changes reduce energy use while others have a limited impact. Gottschalk et al.~\cite{gottschalk2014saving} proposed refactoring-based approaches validated through measurement. Additional work proposed mechanism-level optimizations, including display management~\cite{kim2015content}, GUI recoloring and optimization~\cite{li2015nyx,linares2017gemma}, network request batching~\cite{cai2015delaydroid,pathak2012energy}, UI update batching~\cite{chowdhury2019greenbundle}, and kernel-level improvements~\cite{corral2015energy}. Pinto et al.~\cite{pinto2014mining} and Moura et al.~\cite{moura2015mining} mined StackOverflow posts and GitHub code changes, respectively, identifying recurring energy-saving strategies and common pitfalls, but without focusing on agent-authored or AI-generated artifacts. 

More recently, researchers evaluated the energy efficiency of AI-assisted coding tools. Vartziotis et al.~\cite{vartziotis2024learn} assessed green code generation from tools such as Copilot, ChatGPT-3, and CodeWhisperer. Tuttle et al.~\cite{tuttle2024can} found that LLM-generated code often consumes more energy than human-written code. In contrast, Stivala et al.~\cite{stivala2024investigating} showed that Copilot reduces energy consumption when explicitly prompted, highlighting trade-offs between objectives. 
\emph{However, these works did not examine which energy concerns the LLM-assisted agents address and how exactly they perform energy optimization. This understanding is crucial for making future AI agents more energy aware.}

\section{Methodology}

For this study, we utilized the \textit{all\_pull\_request} subset of the AIDev~\cite{li2025aidev} dataset (Zenodo v3~\cite{li_2025_16919272}), which contains GitHub pull requests authored by five leading AI coding agents: Claude Code, Cursor, Devin, GitHub Copilot, and OpenAI Codex. We did not select the subset of \emph{popular} projects only because that gives us only 41 energy-aware PRs, hindering the robustness of our analysis. For our qualitative analysis, we adopted the thematic analysis approach, which is common to discover themes from text and code~\cite{pinto2014mining, ahmed2025exploring, friesen2025repeat, moura2015mining}. For this, we followed the guidelines of Cruzes \textit{et. al}~\cite{CruzesD.S.2011RSfT}---namely, reading and identifying relevant text segments, initial coding, generating themes and merging them to higher-level themes, and validation.

To identify PRs related to energy consumption, we applied a keyword-based filtering strategy~\cite{pinto2014mining, moura2015mining, bao2016android, cruz2019catalog}. A PR is selected if it contains at least one energy-related and one action-related term, 
where $\textit{EnergyTerm} \in \{\textit{energy, power, battery}\}$ and $\textit{ActionTerm} \in$ \{\textit{consumption, efficiency, saving, optimization, leak, drain, time, profiling}\}~\cite{pinto2014mining, moura2015mining}. To capture different morphological variants of these terms (e.g., \textit{consume/consumed/consumption}, \textit{efficient/efficiency}, \textit{optimize/optimise}), we matched common term stems rather than exact word forms. All matching is performed in a case-insensitive manner on the content (titles and descriptions) of PRs. This filtering step yielded 1,295 candidate PRs.


Keyword-based search, however, can produce many false positives~\cite{pinto2014mining,moura2015mining}. Therefore, we performed a manual evaluation to identify PRs that explicitly address energy consumption. The PR descriptions were written in Markdown and often are very long, making the manual analysis extremely challenging. To alleviate this issue, we developed a visualization tool\footnote{\url{https://github.com/SQMLab/LLM-energy/tree/main/Annotely}} that renders PR content in a format similar to GitHub’s README view, and highlights the hit terms from our provided list, making it much faster to read relevant content. As an initial calibration step, the first and second authors (two grad students in software engineering) jointly labeled 30 PRs as \textit{``Energy''} or \textit{``Not Energy''} in a meeting with the last author, who has 6+ years of experience in software energy research. 

After being trained, the first two authors then independently labeled 100 PRs and computed inter-rater agreement, achieving a Cohen’s kappa of 0.84. Disagreements were discussed and resolved with the third author acting as an adjudicator with 2+ years of industry experience. This process was repeated on another set of 100 PRs, resulting in a kappa of 0.89. After resolving the disagreements, the remaining PRs were divided between the first and second authors for labeling. Finally, all PRs labeled as \textit{``Energy''} were subsequently cross-reviewed by both the first and second authors, and disagreements in 17 PRs were resolved through discussion with the third author. At the end, 216 PRs were labeled as \textit{``Energy''}. 

The first, second, and last authors performed open coding on an initial set of 30 PRs. After the coding process became transparent and easy to apply for everyone, the remaining PRs were then divided such that each of the first two authors coded half of them and reviewed the other half. They then arranged a meeting with the third author to convert all the codes into lower-level and higher-level themes. For example, \emph{efficient data structure/library} is a lower-level theme that, together with other optimization strategies, forms the higher-level \emph{optimization theme}. The fully coded and themed dataset is available with our replication package.

\section{Approach, Analysis \& Results}
This section presents the analysis and results for both RQs.

\subsection{RQ1: Taxonomy of Energy Concerns}

\begin{figure}[htbp]
    \centering
    \includegraphics[width=1\linewidth, keepaspectratio]{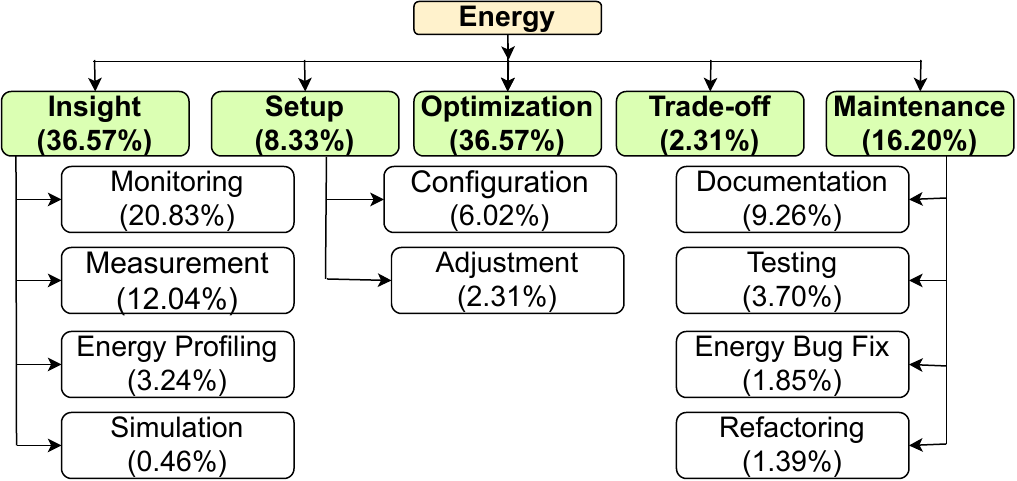}
    \Description{Taxonomy of Energy Concerns.}
    \caption{Taxonomy of Energy Concerns.}
    \label{fig:taxonomy}
\end{figure}

Our thematic analysis yielded five high-level recurring themes/ categories that capture how the agents address energy concerns. Figure~\ref{fig:taxonomy} shows these themes with their sub-themes; trade-off did not have any sub-themes, and the sub-themes for optimization are described in RQ2. Here, we summarize the five higher-level themes, and details are available with the replication package.


\begin{enumerate}[wide, labelwidth=!, labelindent=0pt]
    \item \textbf{Insight:} PRs that make energy behavior observable by tracking, quantifying, profiling hotspots, or simulating energy/power usage.
    \item \textbf{Setup:} PRs that prepare or tune energy-related behavior through settings and parameter/rule updates, without mentioning any energy-saving intent.
    \item \textbf{Optimization:} PRs that explicitly aim to reduce energy, power, or battery usage through efficiency improvements (e.g., reduced computation, batching, memory efficiency, lower FPS).
    \item \textbf{Trade-off:} PRs that mention compromises of energy efficiency to improve other factors, including accuracy, and code quality.
    \item \textbf{Maintenance:} PRs that sustain energy-aware behavior over time via documentation, validation, fixing energy-related defects, or code restructuring for maintainability.
\end{enumerate}

Figure~\ref{fig:taxonomy} shows that AI agents address energy concerns mainly through \textit{Optimization} (36.57\%) and \textit{Insight} (36.57\%). \textit{Insight} is dominated by \textit{Monitoring} (20.83\%) and \textit{Measurement} (12.04\%), indicating a focus on making energy behavior observable. \textit{Maintenance} accounts for 16.20\%, largely via \textit{Documentation} (9.26\%) and \textit{Testing} (3.70\%). The \textit{Trade-off} theme further indicates that agents recognize conflicts between energy efficiency and other attributes, such as code quality and maintainability, as noted in prior work~\cite{chowdhury2019greenbundle, cruz2019energy}.

\begin{figure}[hbtp]
    \centering
    \includegraphics[width=1\linewidth, keepaspectratio]{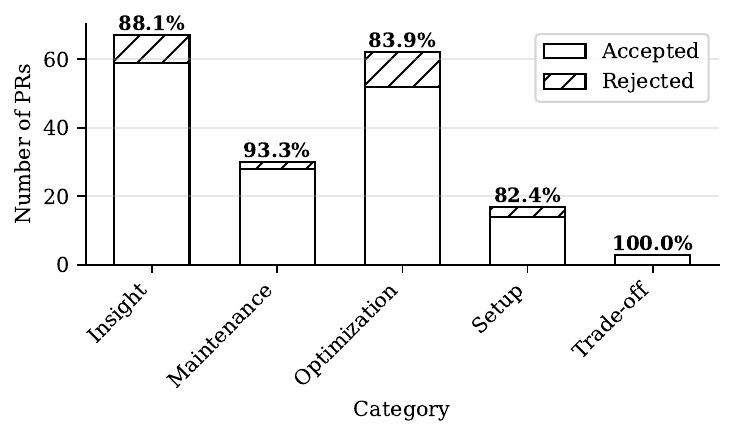}
    \Description{Percent of acceptance/rejection rates of different energy concerns.}
    \caption{Number of accepted/rejected PRs. The numeric values represent the percentage of accepted PRs.}
    \label{fig:accept-reject}
    \vspace{-3mm}
\end{figure}

We found that energy-aware PRs have a slightly lower acceptance rate ($\sim$87\%) than non-energy PRs ($\sim$92\%), and the acceptance rates can vary by category, as shown in Figure~\ref{fig:accept-reject}. To our surprise, the \textit{Optimization} PRs—the most common and arguably most impactful category as they intend to apply direct energy optimization~\cite{pinto2014mining, moura2015mining}—have one of the highest rejection rates, and when accepted, they have the longest review/merge times (merge-time figure is available in the replication package). Both the Wilcoxon rank-sum test and Cliff’s delta---commonly used in SE research~\cite{chowdhury2018exploratory,chowdhury2019greenscaler}---confirm that these observed merge-time differences of the \textit{Optimization} PRs with other categories are statistically significant with small to large (non-negligible) effect sizes.

\begin{figure}[hbtp]
    \centering
    \includegraphics[width=1\linewidth, keepaspectratio]{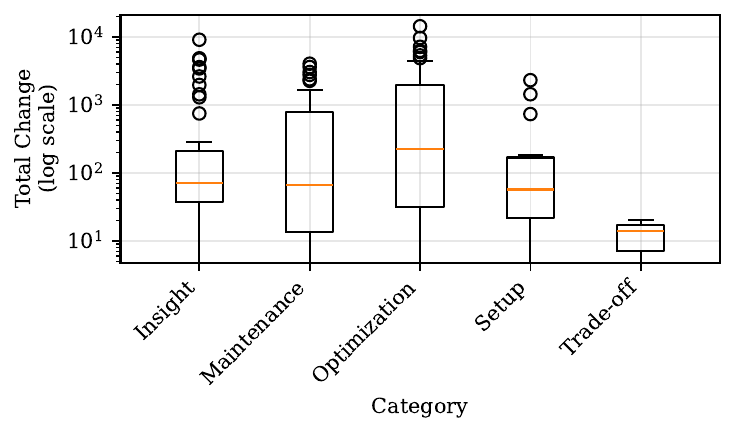}
    \Description{Distribution of change sizes.}
    \caption{Distribution of change sizes.}
    \label{fig:change-size}
\end{figure}

\emph{Why are \textit{Optimization} PRs rejected more often, despite their potential to directly reduce energy consumption?} Cruz et al.~\cite{cruz2019energy} showed that energy optimizations can harm code quality and long-term maintainability. To verify whether this is the case here as well, we opted to compare the change sizes of \textit{Optimization} PRs with others, as change size often correlated with code quality~\cite{chowdhury2019greenscaler}. Unfortunately, change size information was unavailable for all PRs in the AIDev~\cite{li2025aidev} dataset. We therefore used the GitHub API\footnote{\url{https://docs.github.com/en/rest/pulls/pulls?apiVersion=2022-11-28\#list-pull-requests-files}. Script is available with the replication package.} to obtain file-level change sizes for 194 out of 216 PRs; the remaining PRs were inaccessible due to repository deletion or privacy changes. Figure~\ref{fig:change-size} shows that \textit{Optimization} PRs have significantly larger change sizes than other categories, a result confirmed by the aforementioned statistical tests. We also find that \textit{Optimization} PRs modify more files than other categories, indicating higher change complexity (see replication package). Consistent with these observations, in our \textit{Trade-off} category, all five PRs were accepted, where energy efficiency was sacrificed to improve other quality indicators.

\textbox{\textbf{Answer to RQ1.} AI coding agents exhibit energy awareness across multiple dimensions, with the strongest emphasis on \textit{optimization} and observation (\textit{Insight}). However, despite their potential, \textit{optimization} PRs are accepted less frequently due to their negative impact on code quality.}


\subsection{RQ2: Optimization Techniques}
As energy efficiency became an increasingly important software quality concern~\cite{pinto2017energy, muralidhar2022energy}, we aim to support improving future AI-agents by examining how current agents attempt to reduce energy consumption in practice. Assessing whether these techniques actually reduce energy consumption is essential, as intent alone is insufficient. However, direct energy measurement would require a separate full study due to its inherent complexity~\cite{chowdhury2019greenscaler, pinto2014mining}. Instead, we evaluate whether the agents’ intended techniques are supported by prior empirical research that has already measured the energy impact of such techniques.

Based on our thematic analysis, we found 79 PRs that explicitly targeted energy optimizations. We analyzed PR descriptions to extract concrete strategies and mapped them to established prior research on energy optimization. This process yielded 21 distinct optimization techniques (i.e., sub-themes), where each PR employed one or more techniques. Although these 21 sub-themes could be merged into broader themes, we chose to present them at the most possible granular level. To map with existing research, we leveraged two PhD theses on software energy~\cite{da2019tools, chowdhury2019towards} and established survey papers on energy research and practice~\cite{pang2015programmers,georgiou2019software, cruz2019catalog}. Together, these papers summarize most of the known (if not all) energy optimization techniques. Table~\ref{tab:opt_category_distribution} shows all the discovered techniques related to optimization, along with at least one research paper that supported such optimization. We now briefly describe some of these techniques (the replication package contains the rest).

\begin{table}[b]
\centering
\caption{Distribution of agent-applied optimization techniques and their alignment with prior research.}
\label{tab:opt_category_distribution}
\setlength{\tabcolsep}{0pt}
\renewcommand{\arraystretch}{1.15}
\rowcolors{2}{gray!12}{white} 
\begin{tabularx}{\linewidth}{p{0.42\linewidth} >{\centering}p{0.23\linewidth} >{\arraybackslash}p{0.35\linewidth}}

\toprule
\textbf{Optimization Techniques} & \textbf{Frequency} & \makecell[c]{\textbf{Related Work}} \\
\midrule
Avoid unnecessary work & 18 & \cite{moura2015mining, cruz2019catalog, sahin2016benchmarks, li2014investigation}\\
Efficient data structure / library & 10 & \cite{moura2015mining, chowdhury2019greenscaler, pinto2014mining, hasan2016energy, pereira2016influence, manotas2014seeds}\\

Avoid extraneous graphics and animations & 9 & \cite{cruz2019catalog, kim2015content} \\
Decrease rate & 8 & \cite{cruz2019catalog, banerjee2014energy} \\
Power save mode & 6 &  \cite{cruz2019catalog, cruz2019energy}\\
Avoid polling & 5 & \cite{moura2015mining, pinto2014mining, boonkrong2015reducing}\\

Disabling features or devices & 4 & \cite{moura2015mining}\\
Low power idling & 3 & \cite{moura2015mining, metri2012eating}\\
Caching & 3 &  \cite{cruz2019catalog, gottschalk2014saving}\\
Optimizing wake locks & 3 & \cite{bao2016android, liu2014greendroid, liu2016understanding, alam2014energy}\\
Display and UI tuning & 3 &  \cite{moura2015mining, cruz2019catalog, agolli2017investigating, linares2017gemma, li2014investigation, li2015nyx, dong2011chameleon}\\
No screen interaction & 2 &  \cite{cruz2019catalog}\\
Keep IO to a minimum  & 2 &  \cite{pinto2014mining}\\
Reduced resolution   & 2 &  \cite{cruz2019catalog, flinn1999energy}\\
Concurrent programming    & 2 & \cite{pinto2014mining, chowdhury2016greenoracle}\\
Lazy initialization & 2 &  \cite{pinto2014mining}\\
Hardware coordination        & 2 &  \cite{pinto2014mining}\\
\makecell[l]{Dynamic scaling \&\\adaptive throttling}  & 2 & \cite{moura2015mining}\\
Timing out & 1 & \cite{moura2015mining}\\
Batch operations  & 1 &  \cite{cruz2019catalog, li2014investigation, corral2015energy, cai2015delaydroid, chowdhury2019greenbundle, chowdhury2018exploratory}\\
\midrule
\multicolumn{3}{l}{\textbf{Other Techniques}} \\
\midrule
Not specified & 13 & --- \\
Optimized sampling    & 2 & ---\\
 
\bottomrule
\end{tabularx}
\end{table}

\begin{itemize}[wide, labelwidth=!, labelindent=0pt]
 
\item \textbf{Avoid unnecessary work:} These techniques reduce energy by removing redundant computations, unnecessary re-renders, repeated checks, or duplicated logic.

\item \textbf{Efficient data structure/library:} These techniques reduce energy consumption by replacing an expensive data structure or library with a more energy-efficient one.

\item \textbf{Decrease rate:} These techniques save energy by reducing how often updates happen (e.g., events, rendering, syncing, and refresh).

\item \textbf{Avoid polling:} These reduce energy by replacing frequent periodic checks with event-driven, conditional, or less frequent updates (i.e., pushing rather than pulling). 

\item \textbf{Caching:} These techniques store computed results to avoid repeated expensive operations.

\item \textbf{Optimizing wake locks:} Mainly in Android systems, these techniques keep the wake locks only when needed and release them immediately after the work is done.

\item \textbf{Concurrent programming:} These reduce active time or avoid blocking, enabling faster completion and earlier idle states. 

\item \textbf{Hardware coordination:} Techniques to coordinate hardware components or sensors to avoid conflicts and unnecessary use.

\item \textbf{Dynamic scaling \& adaptive throttling:} 
Dynamically adjusts quality, frequency, or workload based on runtime context (e.g., device capability, system load, visibility, and thermal state), including selecting different throttle intervals or performance tiers.

\item \textbf{Batch operations:} Groups multiple operations (e.g., bundle view updates, bundled network operations) to reduce repeated overhead, wake-ups, and tail energy leaks.

\item \textbf{Not specified:} PRs that claim reduced energy consumption but do not explain what specific techniques are applied.

\item \textbf{Optimized sampling:} Lowers energy use by collecting or processing data less often (or only when needed).

\end{itemize}

In general, instead of complex redesign, the AI agents most often apply direct workload-reduction strategies, such as \textit{``avoid unnecessary work''}, \textit{``avoid extraneous graphics and animations''}, \textit{``Decrease rate''}, and \textit{``Avoid polling''} suggesting, the agents primarily target clear CPU/GPU overhead such as redundant computation, repeated UI work, and non-essential visual effects. Encouragingly, most of these techniques (20 out of 21) align with established empirically validated recommendations.



\textbox{\textbf{Answer to RQ2.} AI agents primarily rely on empirically validated established energy optimization techniques—such as reducing redundant work and graphical overhead, adapting execution frequency, and using efficient data structures—while providing very limited novel energy-saving strategies.} \vspace{-0.4cm}

\section{Threats to validity}
Several threats can impact the validity of our findings.

\textbf{Internal validity.} Our keyword-based filtering and manual validation may miss PRs that address energy concerns implicitly. Also, brief or incomplete PR descriptions can limit accurate interpretation of intent or applied techniques. Moreover, thematic analysis involves subjective judgment and bias. This threat was mitigated by multiple rounds of independent labeling and discussion about disagreements. 

\textbf{External validity.} The dataset we used contains five AI coding agents, which may not fully represent other agents, tools, or all forms of energy-aware agent behavior. A staggering 87\% of the PRs are from the \textit{OpenAI Codex} agent, which is a limitation of the used dataset and our study.
\section{Conclusion}

In this paper, we analyzed agent-authored pull requests to understand how modern AI coding agents address energy concerns, which energy optimization techniques they apply, and whether these techniques are supported by research findings. We identified five broader themes of energy concerns considered by AI agents and 21 distinct energy optimization techniques, many of which are empirically validated by prior studies. Our results suggest that future research should focus on techniques that reduce energy consumption without harming other quality attributes, as this could improve the acceptance of energy-aware pull requests.

While these findings are generally encouraging, future work should focus on empirically validating the code changes implemented by these agents to ensure that intended energy optimizations are accurately realized in practice. We hope our findings will inspire the research community to train future AI agents to be even more energy-conscious, ultimately supporting the creation of more sustainable software systems. 

\bibliographystyle{ACM-Reference-Format}
\bibliography{sample-base, acmart}

@article{li2025aidev,
title={{The Rise of AI Teammates in Software Engineering (SE) 3.0: How Autonomous Coding Agents Are Reshaping Software Engineering}}, 
author={Li, Hao and Zhang, Haoxiang and Hassan, Ahmed E.},
journal={arXiv preprint arXiv:2507.15003},
year={2025}
}

@dataset{li_2025_16919272,
  author       = {Li, Hao and
                  Zhang, Haoxiang and
                  Hassan, Ahmed E.},
  title        = {AIDev: Studying AI Coding Agents on GitHub},
  month        = nov,
  year         = 2025,
  publisher    = {Zenodo},
  doi          = {10.5281/zenodo.16919272},
  url          = {https://doi.org/10.5281/zenodo.16919272},
}

@article{koot2021usage,
  title={Usage impact on data center electricity needs: A system dynamic forecasting model},
  author={Koot, Martijn and Wijnhoven, Fons},
  journal={Applied Energy},
  volume={291},
  pages={116798},
  year={2021},
  publisher={Elsevier}
}

@article{faiz2023llmcarbon,
  title={Llmcarbon: Modeling the end-to-end carbon footprint of large language models},
  author={Faiz, Ahmad and Kaneda, Sotaro and Wang, Ruhan and Osi, Rita and Sharma, Prateek and Chen, Fan and Jiang, Lei},
  journal={arXiv preprint arXiv:2309.14393},
  year={2023}
}

@article{argerich2024measuring,
  title={Measuring and improving the energy efficiency of large language models inference},
  author={Argerich, Mauricio Fadel and Pati{\~n}o-Mart{\'\i}nez, Marta},
  journal={IEEE Access},
  volume={12},
  pages={80194--80207},
  year={2024},
  publisher={IEEE}
}

@misc{user-care-battery,
  title = {customers really want better battery life},
  author = {V. Woollaston},
  year = {2014},
  url = {http:// www.dailymail.co.uk/sciencetech/article-2715860/},
  note = {last accessed: 2025-Dec-20}
}

@inproceedings{chowdhury2019greenbundle,
  title={Greenbundle: an empirical study on the energy impact of bundled processing},
  author={Chowdhury, Shaiful Alam and Hindle, Abram and Kazman, Rick and Shuto, Takumi and Matsui, Ken and Kamei, Yasutaka},
  booktitle={2019 41st ICSE},
  pages={1107--1118},
  year={2019},
  organization={IEEE}
}

@inproceedings{moura2015mining,
  title={Mining energy-aware commits},
  author={Moura, Irineu and Pinto, Gustavo and Ebert, Felipe and Castor, Fernando},
  booktitle={2015 IEEE/ACM 12th Working Conference on Mining Software Repositories},
  pages={56--67},
  year={2015},
  organization={IEEE}
}

@inproceedings{osta2019energy,
  title={Energy efficient implementation of machine learning algorithms on hardware platforms},
  author={Osta, Mario and Alameh, Mohamad and Younes, Hamoud and Ibrahim, Ali and Valle, Maurizio},
  booktitle={2019 26th ICECS},
  pages={21--24},
  year={2019},
  organization={IEEE}
}

@inproceedings{CruzesD.S.2011RSfT,
author = {Cruzes, D. S. and Dyba, T.},
booktitle = {2011 ESEM},
pages = {275-284},
title = {Recommended Steps for Thematic Synthesis in Software Engineering},
year = {2011},
}

@inproceedings{carroll2010analysis,
  title={An analysis of power consumption in a smartphone},
  author={Carroll, Aaron and Heiser, Gernot},
  booktitle={2010 USENIX Annual Technical Conference (USENIX ATC 10)},
  year={2010}
}

@article{hassan2024towards,
  title={Towards AI-native software engineering (SE 3.0): A vision and a challenge roadmap},
  author={Hassan, Ahmed E and Oliva, Gustavo A and Lin, Dayi and Chen, Boyuan and Ming, Zhen and others},
  journal={arXiv preprint arXiv:2410.06107},
  year={2024}
}

@inproceedings{cursaru2024controlled,
  title={A controlled experiment on the energy efficiency of the source code generated by code llama},
  author={Cursaru, Vlad-Andrei and Duits, Laura and Milligan, Joel and Ural, Damla and Sanchez, Berta Rodriguez and Stoico, Vincenzo and Malavolta, Ivano},
  booktitle={QUATIC},
  pages={161--176},
  year={2024},
  organization={Springer}
}

@article{lee2024survey,
  title={A survey of energy concerns for software engineering},
  author={Lee, Sung Une and Fernando, Niroshinie and Lee, Kevin and Schneider, Jean-Guy},
  journal={Journal of Systems and Software},
  volume={210},
  pages={111944},
  year={2024},
  publisher={Elsevier}
}

@article{cruz2019catalog,
  title={Catalog of energy patterns for mobile applications},
  author={Cruz, Luis and Abreu, Rui},
  journal={Empirical software engineering},
  volume={24},
  number={4},
  pages={2209--2235},
  year={2019},
  publisher={Springer}
}

@inproceedings{bao2016android,
  title={How Android app developers manage power consumption? An empirical study by mining power management commits},
  author={Bao, Lingfeng and Lo, David and Xia, Xin and Wang, Xinyu and Tian, Cong},
  booktitle={In 13th MSR},
  pages={37--48},
  year={2016}
}

@inproceedings{pinto2014mining,
  title={Mining questions about software energy consumption},
  author={Pinto, Gustavo and Castor, Fernando and Liu, Yu David},
  booktitle={Proceedings of the 11th working conference on mining software repositories},
  pages={22--31},
  year={2014}
}

@inproceedings{agolli2017investigating,
  title={Investigating decreasing energy usage in mobile apps via indistinguishable color changes},
  author={Agolli, Tedis and Pollock, Lori and Clause, James},
  booktitle={2017 IEEE/ACM 4th  MOBILESoft},
  pages={30--34},
  year={2017},
  organization={IEEE}
}

@inproceedings{linares2017gemma,
  title={Gemma: multi-objective optimization of energy consumption of guis in android apps},
  author={Linares-V{\'a}squez, Mario and Bernal-C{\'a}rdenas, Carlos and Bavota, Gabriele and Oliveto, Rocco and Di Penta, Massimiliano and Poshyvanyk, Denys},
  booktitle={2017 IEEE/ACM 39th International Conference on Software Engineering Companion},
  pages={11--14},
  year={2017},
  organization={IEEE}
}

@inproceedings{li2014investigation,
  title={An investigation into energy-saving programming practices for android smartphone app development},
  author={Li, Ding and Halfond, William GJ},
  booktitle={Proceedings of the 3rd International Workshop on Green and Sustainable Software},
  pages={46--53},
  year={2014}
}

@inproceedings{li2015nyx,
  title={Nyx: A display energy optimizer for mobile web apps},
  author={Li, Ding and Tran, Angelica Huyen and Halfond, William GJ},
  booktitle={Proceedings of the 2015 10th Joint Meeting on Foundations of Software Engineering},
  pages={958--961},
  year={2015}
}

@inproceedings{corral2015energy,
  title={Energy-aware performance evaluation of android custom kernels},
  author={Corral, Luis and Georgiev, Anton B and Janes, Andrea and Kofler, Stefan},
  booktitle={2015 IEEE/ACM 4th International Workshop on Green and Sustainable Software},
  pages={1--7},
  year={2015},
  organization={IEEE}
}

@inproceedings{cai2015delaydroid,
  title={Delaydroid: Reducing tail-time energy by refactoring android apps},
  author={Cai, Huaqian and Zhang, Ying and Jin, Zhi and Liu, Xuanzhe and Huang, Gang},
  booktitle={Proceedings of the 7th Asia-Pacific Symposium on Internetware},
  pages={1--10},
  year={2015}
}

@article{kim2015content,
  title={Content-centric energy management of mobile displays},
  author={Kim, Dongwon and Jung, Nohyun and Chon, Yohan and Cha, Hojung},
  journal={IEEE Transactions on Mobile Computing},
  volume={15},
  number={8},
  pages={1925--1938},
  year={2015},
  publisher={IEEE}
}

@inproceedings{gottschalk2014saving,
  title={Saving energy on mobile devices by refactoring.},
  author={Gottschalk, Marion and Jelschen, Jan and Winter, Andreas},
  booktitle={EnviroInfo},
  pages={437--444},
  year={2014}
}

@inproceedings{vartziotis2024learn,
  title={Learn to code sustainably: An empirical study on green code generation},
  author={Vartziotis, Tina and Dellatolas, Ippolyti and Dasoulas, George and Schmidt, Maximilian and Schneider, Florian and Hoffmann, Tim and Kotsopoulos, Sotirios and Keckeisen, Michael},
  booktitle={Proceedings of the 1st International Workshop on Large Language Models for Code},
  pages={30--37},
  year={2024}
}

@incollection{stivala2024investigating,
  title={Investigating the Use of GitHub Copilot for Green Software},
  author={Stivala, Maria and Fatima, Iffat and Lago, Patricia},
  booktitle={Environmental Informatics},
  pages={219--235},
  year={2024},
  publisher={Springer}
}

@inproceedings{tuttle2024can,
  title={Can llms generate green code-a comprehensive study through leetcode},
  author={Tuttle, Jonas F and Chen, Dayuan and Nasrin, Amina and Soto, Noe and Zong, Ziliang},
  booktitle={2024 IEEE 15th IGSC},
  pages={39--44},
  year={2024},
  organization={IEEE}
}

@article{pinto2017energy,
  title={Energy efficiency: a new concern for application software developers},
  author={Pinto, Gustavo and Castor, Fernando},
  journal={Communications of the ACM},
  volume={60},
  number={12},
  pages={68--75},
  year={2017},
  publisher={ACM New York, NY, USA}
}

@article{muralidhar2022energy,
  title={Energy efficient computing systems: Architectures, abstractions and modeling to techniques and standards},
  author={Muralidhar, Rajeev and Borovica-Gajic, Renata and Buyya, Rajkumar},
  journal={ACM Computing Surveys (CSUR)},
  volume={54},
  number={11s},
  pages={1--37},
  year={2022},
  publisher={ACM New York, NY}
}

@inproceedings{hasan2016energy,
  title={Energy profiles of java collections classes},
  author={Hasan, Samir and King, Zachary and Hafiz, Munawar and Sayagh, Mohammed and Adams, Bram and Hindle, Abram},
  booktitle={Proceedings of the 38th ICSE},
  pages={225--236},
  year={2016}
}

@inproceedings{pereira2016influence,
  title={The influence of the java collection framework on overall energy consumption},
  author={Pereira, Rui and Couto, Marco and Saraiva, Jo{\~a}o and Cunha, J{\'a}come and Fernandes, Jo{\~a}o Paulo},
  booktitle={Proceedings of the 5th International Workshop on Green and Sustainable Software},
  pages={15--21},
  year={2016}
}

@inproceedings{manotas2014seeds,
  title={Seeds: A software engineer's energy-optimization decision support framework},
  author={Manotas, Irene and Pollock, Lori and Clause, James},
  booktitle={In the 36th ICSE},
  pages={503--514},
  year={2014}
}

@article{chowdhury2018exploratory,
  title={An exploratory study on assessing the energy impact of logging on android applications},
  author={Chowdhury, Shaiful and Di Nardo, Silvia and Hindle, Abram and Jiang, Zhen Ming},
  journal={Empirical Software Engineering},
  volume={23},
  number={3},
  pages={1422--1456},
  year={2018},
  publisher={Springer}
}

@article{chowdhury2019greenscaler,
  title={Greenscaler: training software energy models with automatic test generation},
  author={Chowdhury, Shaiful and Borle, Stephanie and Romansky, Stephen and Hindle, Abram},
  journal={Empirical Software Engineering},
  volume={24},
  number={4},
  pages={1649--1692},
  year={2019},
  publisher={Springer}
}

@inproceedings{chowdhury2016greenoracle,
  title={Greenoracle: Estimating software energy consumption with energy measurement corpora},
  author={Chowdhury, Shaiful Alam and Hindle, Abram},
  booktitle={Proceedings of the 13th international conference on mining software repositories},
  pages={49--60},
  year={2016}
}

@article{sahin2016benchmarks,
  title={From benchmarks to real apps: Exploring the energy impacts of performance-directed changes},
  author={Sahin, Cagri and Pollock, Lori and Clause, James},
  journal={Journal of Systems and Software},
  volume={117},
  pages={307--316},
  year={2016},
  publisher={Elsevier}
}

@inproceedings{dong2011chameleon,
  title={Chameleon: A color-adaptive web browser for mobile OLED displays},
  author={Dong, Mian and Zhong, Lin},
  booktitle={Proceedings of the 9th international conference on Mobile systems, applications, and services},
  pages={85--98},
  year={2011}
}

@inproceedings{boonkrong2015reducing,
  title={Reducing battery consumption of data polling and pushing techniques on android using gzip},
  author={Boonkrong, Sirapat and Dinh, Pham Cao},
  booktitle={2015 7th ICITEE},
  pages={565--570},
  year={2015},
  organization={IEEE}
}

@inproceedings{metri2012eating,
  title={What is eating up battery life on my SmartPhone: A case study},
  author={Metri, Grace and Agrawal, Abhishek and Peri, Ramesh and Shi, Weisong},
  booktitle={2012 International Conference on Energy Aware Computing},
  pages={1--6},
  year={2012},
  organization={IEEE}
}

@inproceedings{liu2016understanding,
  title={Understanding and detecting wake lock misuses for android applications},
  author={Liu, Yepang and Xu, Chang and Cheung, Shing-Chi and Terragni, Valerio},
  booktitle={Proceedings of the 2016 24th FSE},
  pages={396--409},
  year={2016}
}

@article{liu2014greendroid,
  title={Greendroid: Automated diagnosis of energy inefficiency for smartphone applications},
  author={Liu, Yepang and Xu, Chang and Cheung, Shing-Chi and L{\"u}, Jian},
  journal={IEEE Transactions on Software Engineering},
  volume={40},
  number={9},
  pages={911--940},
  year={2014},
  publisher={IEEE}
}

@inproceedings{banerjee2014energy,
  title={Energy-aware design patterns for mobile application development (invited talk)},
  author={Banerjee, Abhijeet and Roychoudhury, Abhik},
  booktitle={Proceedings of the 2Nd International Workshop on Software Development Lifecycle for Mobile},
  pages={15--16},
  year={2014}
}

@inproceedings{alam2014energy,
  title={Energy optimization in android applications through wakelock placement},
  author={Alam, Faisal and Panda, Preeti Ranjan and Tripathi, Nikhil and Sharma, Namita and Narayan, Sanjiv},
  booktitle={2014 Design, DATE},
  pages={1--4},
  year={2014},
  organization={IEEE}
}

@article{flinn1999energy,
  title={Energy-aware adaptation for mobile applications},
  author={Flinn, Jason and Satyanarayanan, Mahadev},
  journal={ACM SIGOPS Operating Systems Review},
  volume={33},
  number={5},
  pages={48--63},
  year={1999},
  publisher={ACM New York, NY, USA}
}

@String{Computing = "Computing" }

@String{Computer = "{IEEE} Computer" }

@String{Springer = "Springer-Verlag" }

@article{friesen2025repeat,
  title={The Repeat Offenders: Characterizing and Predicting Extremely Bug-Prone Source Methods},
  author={Friesen, Ethan and Morton-Salmon, Sasha and Opu, Md Nahidul Islam and Islam, Shahidul and Chowdhury, Shaiful},
  journal={arXiv preprint arXiv:2511.22726},
  year={2025}
}

@phdthesis{da2019tools,
  title={Tools and techniques for energy-efficient mobile application development},
  author={da Cruz, Lu{\'\i}s Miranda},
  year={2019},
  school={Universidade do Porto (Portugal)}
}

@inproceedings{cruz2019energy,
  title={Do energy-oriented changes hinder maintainability?},
  author={Cruz, Luis and Abreu, Rui and Grundy, John and Li, Li and Xia, Xin},
  booktitle={2019 ICSME},
  year={2019}
}

@article{georgiou2019software,
  title={Software development lifecycle for energy efficiency: techniques and tools},
  author={Georgiou, Stefanos and Rizou, Stamatia and Spinellis, Diomidis},
  journal={ACM Computing Surveys (CSUR)},
  volume={52},
  number={4},
  year={2019},
 
}

@article{pang2015programmers,
  title={What do programmers know about software energy consumption?},
  author={Pang, Candy and Hindle, Abram and Adams, Bram and Hassan, Ahmed E},
  journal={IEEE Software},
  volume={33},
  number={3},
  year={2015},
}

@article{chowdhury2019towards,
  title={Towards Developing Energy Efficient Mobile Applications: Models, Tools, and Guidelines},
  author={Chowdhury, Shaiful Alam},
  year={2019},
 school={University of Alberta (Canada)}
}

@article{ahmed2025exploring,
  title={Exploring Challenges in Test Mocking: Developer Questions and Insights from StackOverflow},
  author={Ahmed, Mumtahina and Opu, Md Nahidul Islam and Roy, Chanchal and Suhi, Sujana Islam and Chowdhury, Shaiful},
  journal={arXiv preprint arXiv:2505.08300},
  year={2025}
}

@inproceedings{pathak2012energy,
  title={Where is the energy spent inside my app? Fine Grained Energy Accounting on Smartphones with Eprof},
  author={Pathak, Abhinav and Hu, Y Charlie and Zhang, Ming},
  booktitle={Proceedings of the 7th ACM european conference on Computer Systems},
  pages={29--42},
  year={2012}
}

\end{document}